\documentclass[aps,prl,twocolumn,amsmath,amssymb,amsfonts,superscriptaddress]{revtex4}

\usepackage{graphicx}
\usepackage{subfigure}
\usepackage{times} 

\newcommand{\avg}[1]{\langle #1 \rangle}

\newcommand{\h}[1]{{#1}^{\dagger}} 
\newcommand{\cc}[1]{{#1}^{*}}
\newcommand{\cb}[1]{\bar{#1}}

\newcommand{\up}{\uparrow}
\newcommand{\down}{\downarrow}
\newcommand{\bk}{\vec{k}}

\newcommand{\del} {\partial}

\newcommand{\dmit}{{[{\rm Pd}({\rm dmit})_2]_2}}
\newcommand{\EtMeP}{{\rm EtMe}_3{\rm P}}
\newcommand{\MeP}{{\rm Me}_4{\rm P}}
\newcommand{\MeSb}{{\rm Me}_4{\rm Sb}}
\newcommand{\MeAs}{{\rm Me}_4{\rm As}}
\newcommand{\EtEtMeP}{{\rm Et}_2{\rm Me}_2{\rm P}}
\newcommand{\EtMeAs}{{\rm Et}_2{\rm Me}_2{\rm As}}
\newcommand{\EtMeSb}{{\rm EtMe}_3{\rm Sb}}

\begin{document}


\title{Emergence of superconductivity, valence bond order and Mott insulators in Pd[(dmit)$_2$] based
organic salts } 
\author{Jeffrey G. Rau}
\affiliation{Department of Physics, University of Toronto, Toronto, Ontario M5S 1A7, Canada}
\author{Hae-Young Kee}
\email[Electronic Address: ]{hykee@physics.utoronto.ca}
\affiliation{Department of Physics, University of Toronto, Toronto, Ontario M5S 1A7, Canada}
\affiliation{Canadian Institute for Advanced Research/Quantum Materials Program, Toronto, Ontario MSG 1Z8, Canada}

\date{\today}

\begin{abstract}
The $\EtMeP$ and $\EtMeSb$ nearly triangular organic salts
are distinguished from most other Pd[(dmit)$_2$] based salts, as they display valence bond and no long range order, respectively. 
Under pressure, a superconducting phase is revealed in EtMe$_3$P near the boundary of valence bond order.
We use slave-rotor theory with an enlarged unit cell to study competition between uniform
and broken translational symmetry states, offering a theoretical framework capturing the superconducting, valence bond order,
spin liquid, and metallic phases on an isotropic triangular lattice. 
Our finite temperature phase diagram manifests a remarkable resemblance
to the phase diagram of the EtMe$_3$P salt, where the re-entrant transitions of the type
 insulator-metal-insulator can be explained by an entropy difference between metal and the U(1) spin liquid.
We find that the superconducting pairing symmetry is $d \pm id$, and predict different temperature dependences of the
specific heat between the spin liquid and metal.
 
\end{abstract}


\maketitle

Quasi two-dimensional organic salts have provided a playground to study the rich physics of frustrated and
strongly correlated systems. One fascinating example is the family of triangular  Z$\dmit$ salts
where  Z is a monovalent cation and dmit refers to 1,3-dithiol-2-thione-4,5-dithiolate.
At ambient pressure, most  $\dmit$ compounds are paramagnetic Mott insulators 
with spin susceptibility showing the characteristic temperature dependence of 
a triangular spin-$1/2$ Heisenberg antiferromagnet at high temperatures \cite{et-review}. 
If geometrical frustration is less effective, one expects long range order to develop
at low temperatures to release the spin entropy. Antiferromagnetic (AF) order is a natural candidate, which
is indeed observed in the cases of Z= $\MeP$, $\MeSb$, $\MeAs$, $\EtEtMeP$ and $\EtMeAs$ \cite{et-review}. 
However, no long range order has been detected in $\EtMeSb$ \cite{number-61}, potentially exhibiting a spin liquid phase, 
and in $\EtMeP$ \cite{et-2006,et-2007} a spin-gapped phase with columnar valence bond (VB) order emerges.

The degree of frustration plays an essential role in the variety of ground states of these materials, despite
the common insulating behavior.  In-plane spatial anisotropy, such as in the hopping or
 Heisenberg exchange integrals, is a way to control the geometrical frustration.
It turns out that the AF order is found in most anisotropic materials listed above,
while for $\EtMeP$ and $\EtMeSb$, where AF order is absent, the lattice is nearly isotropic triangular \cite{et-2006,jacs-2006,et-review}.  

While most $\dmit$ salts become metallic \cite{et-2007} under pressure, 
the $\EtMeP$ salt exhibits superconductivity with maximum $T_c$ appearing
at the border of the VB phase seen in X-ray measurements \cite{et-2006,et-2007}. 
Re-entrant transitions of the form insulator-metal-insulator near the phase boundary at low temperatures were reported \cite{et-2007}, where the positive
curvature of the phase boundary indicates a large entropy contained in the paramagnetic insulator relative
to the metallic phase. 
The complexity presented in the phase diagram of $\EtMeP$ has not yet been explained within 
a single theoretical framework.

In this paper, we provide a microscopic theory which demonstrates the emergence of the superconducting, VB ordered, spin liquid,
and metallic phases. We focus on an isotropic triangular lattice aiming to understand the phenomena observed in the
$\EtMeP$ and $\EtMeSb$ salts. In addition to the well-known on-site Hubbard interaction, we find a spin-spin interaction
plays an essential role in favoring the different ground states reported for $\EtMeP$ and $\EtMeSb$.
Our theory also provides a spin liquid surviving between the metal and VB ordered phases, superconducting
pairing symmetry, and novel behavior of the specific heat across the metal-insulator transition.

A common theoretical approach for Mott insulators is to start from a localized
Heisenberg spin model, which is  appropriate for insulators.
However, $\EtMeP$ becomes metallic/superconducting when pressure is applied, exhibiting a metal-insulator (MI) transition. 
Capturing the MI transition requires an electronic description, such as a Hubbard model at an intermediate coupling \cite{hubbard-heisenberg-2,pirg-1,hubbard-2,hubbard-3}. 
Can one access the competition between spin liquid and VB ordered insulators or between metal and superconductor within the Hubbard model near the MI transition?
Prior work \cite{honerkamp-1,clay-ed} suggests difficulty in achieving a superconducting phase within the pure Hubbard model at intermediate coupling 
in contrast to when a spin-spin interaction is included.
Combining the experimental facts and the previous theoretical work,
we suggest the following $t$-$J$-$U$ Hamiltonian is a minimal model for isotropic triangular $\dmit$ salts.
\begin{equation}
  \label{tuj-model}
  H = - t \sum_{\avg{ij}\sigma} \h{c}_{i\sigma}c_{j\sigma} + U \sum_{i} n_{i\up}n_{i\down} + J \sum_{\avg{ij}} \vec{S}_i \cdot \vec{S}_j,
\end{equation}
where $\sum_{\avg{ij}}$ runs over nearest neighbors, $t>0$ is the nearest neighbor hopping (we will set $t=1$
as our energy scale below), $U$ 
is the on-site Hubbard repulsion, and $J$ is the spin-spin interaction. We emphasize that the $J$ term is being considered as an independent effective interaction.
The $c_{i\sigma}$ and $\h{c}_{i\sigma}$ are the electron operators and the electron spin operator is $\vec{S}_i = \sum_{\alpha \beta} \h{c}_{i\alpha}(\vec{\sigma})_{\alpha \beta} c_{i\beta}$, where $\vec{\sigma}$ are the Pauli matrices.  

We take a slave rotor approach first developed by Florens and Georges \cite{rotor-florens-anderson, rotor-florens-mott} 
which can handle the intermediate coupling regime where charge fluctuations are important.  
In the slave-rotor representation \cite{rotor-florens-mott}, the electron operators are expressed as $c_{i \sigma} = f_{i \sigma} e^{i \theta_i}$,
where $\exp\left(-i \theta_i\right)$ is the rotor lowering operator and $f_{i\sigma}$, $\h{f}_{i\sigma}$ are fermionic spinon operators. Using this
representation  the Hamiltonian (\ref{tuj-model}) is written as \cite{rotor-florens-mott} 
\begin{eqnarray*}
H  & = &
  -t \sum_{\avg{ij}\sigma} \h{f}_{i\sigma}f_{j\sigma} e^{-i (\theta_i -\theta_j)} \\
& &+ \frac{U}{2} \sum_i L_i (L_i -1)+ J \sum_{\avg{ij}} \vec{S}^f_i \cdot \vec{S}^f_j,
\end{eqnarray*}
where $\vec{S}^f_i = 1/2 \sum_{\alpha\beta} \h{f}_{i\alpha}(\vec{\sigma})_{\alpha\beta} f_{i\beta}$ is the spinon spin operator and $L_i$ is the
rotor charge operator \cite{rotor-florens-mott}. It is important that the constraint of $L_i + \sum_{\alpha} f_{i \alpha}^\dagger
f_{i \alpha} = 1$ be satisfied to project the Hilbert space into the physical subspace at half-filling.

\begin{figure}
\begin{tabular}{cc}
\subfigure[]{
 \label{unit-cell-fig}
\parbox[c]{11em}{\includegraphics[scale=0.25]{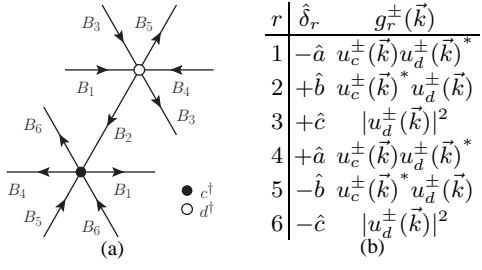}}
}
&
\subfigure[]{
\label{form-factors}
\begin{tabular}{c|cc}
$r$ & $\hat{\delta}_r$ & $g^{\pm}_{r}(\bk)$ \\
\hline
$1$ & $-\hat{a}$ & $u^{\pm}_{c}(\bk)\cc{u^{\pm}_{d}(\bk)}$ \\
$2$ &$+\hat{b}$ & $\cc{u^{\pm}_{c}(\bk)}u^{\pm}_{d}(\bk)$ \\
$3$ & $+\hat{c}$ & $|u^{\pm}_{d}(\bk)|^2$ \\
$4$ & $+\hat{a}$ & $u^{\pm}_{c}(\bk)\cc{u^{\pm}_{d}(\bk)}$ \\
$5$ &$-\hat{b}$ & $\cc{u^{\pm}_{c}(\bk)}u^{\pm}_{d}(\bk)$ \\
$6$ &$-\hat{c}$ & $|u^{\pm}_{d}(\bk)|^2$ \\

\end{tabular}
}
\end{tabular}
\caption{ (a) The doubled unit cell (b) Form factors used in the mean
field equations.
}
\end{figure}

We adopt  the slave rotor theory and extend to a enlarged, two site unit cell to study the competition between 
uniform states and  states which minimally break physical translational invariance.
The spinon operators for the two sites in our unit cell are denoted by $c_{i\sigma}$
and $d_{i,\sigma}$ as indicated in Fig. \ref{unit-cell-fig}.   
Introducing the mean fields, $\chi_{ij} = \frac{1}{2} \sum_{\sigma}\avg{\h{f}_{i\sigma}f_{j\sigma}}$, 
$B_{ij} =  \avg{e^{-i (\theta_i-\theta_j)}}$, and $\Delta_{ij} = \avg{\h{f}_{i\uparrow} \h{f}_{j\downarrow}}$,
the spinon part of the Hamiltonian is given by
\begin{eqnarray*}
H_f & = & - \sum_{\sigma\avg{ij}} \left(tB_{ij} + \frac{3J}{2}\cc{\chi}_{ij}\right)\h{f}_{i\sigma} f_{j\sigma} \\
& -&\frac{3J}{8}\sum_{\avg{ij}}\left(\Delta_{ij} (f_{i\downarrow}f_{j\uparrow} - f_{i\uparrow}f_{j\downarrow}) + {\rm h.c} \right) \\
&+& \sum_{ij} \left(\frac{3J}{2}|\chi_{ij}|^2 + \frac{3J}{8}|\Delta_{ij}|^2\right),
\\
  & \equiv &  \sum_{\bf k} \h{\Psi}_{\bf k} M_{\bf k} \Psi_{\bf k} 
 + N \sum_{r=1}^{6}\left(\frac{3J}{2}|\chi_r|^2 + \frac{3J}{8}|\Delta_r|^2\right) + E_{0}.
\end{eqnarray*}
where $M_{\bf k}$ is $4 \times 4$ matrix in the basis of $\Psi_{\bf k} = 
\left( c^{\dagger}_{{\bf k} \uparrow}, d^{\dagger}_{{\bf k} \uparrow}
 c_{-{\bf k} \downarrow}, d_{-{\bf k} \downarrow} \right)$ reflecting the unit cell doubling in particle-hole
Nambu basis. The additive constant is the Fermi sea energy.
Treating the rotors as a bosonic variable, i.e., $e^{i \theta_i} = \phi_i$ subject to the constraint of $|\phi_i|^2=1$, 
the rotor part of the Hamiltonian can be represented by the (imaginary time) action,
\begin{eqnarray*}
S & = & \int_0^{\beta} d\tau \left(
\sum_i \left(\frac{|\del_\tau\phi_i|^2}{2U}+i\lambda_i|\phi_i|^2\right)
-2t\sum_{\avg{ij}} \chi_{ij} \cb{\phi_i}\phi_j
\right), \\
 & \equiv & \sum_{ k \omega}\left(
\gamma^{+}_k |\zeta_{+}(\bk,\omega)|^2+
\gamma^{-}_k |\zeta_{-}(\bk,\omega)|^2
\right),
\end{eqnarray*}
where the final form is obtained as in the spinon case by decomposing into
the two sublattices and Fourier transforming.
The $i \lambda_i (\equiv h)$ enforces the constraint and acts as a chemical potential.
To consider condensation, we separate out the lowest energy component of the lower
band $\gamma^{-}_k$ by diagonalizing the above action, and assuming $\bk=0$ condensation
can be  treated as an independent variable. 
Denoting the condensate density as $|z|^2$, the saddle point conditions for $h_a$, $h_b$ and
$z$ give constraint equations for the number of bosons.
 
The full set of mean field equations is given as
\begin{eqnarray}
\chi_{r}  & = & 
\frac{1}{N} 
\sum_{k} \sum_{l=1}^4  n_f(E_{l}(\bk)) [\h{U}_k O_{\chi_r}(\bk) U_k]_{ll}  ,\nonumber\\
\Delta_{r}  & = & 
\frac{1}{N} 
\sum_{k }\sum_{l=1}^4  n_f(E_{l}(\bk)) [\h{U}_k O_{\Delta_r}(\bk) U_k]_{ll} ,\nonumber\\
B_{r}  & = & 
\frac{1}{N} 
\sum_{k \alpha} e^{i \bk \cdot \hat{\delta}_r} g^{\alpha}_{r}(k) n_b ( \gamma^{\alpha}_k) 
+  g^{-}_{r}(0)|z|^2,\label{mf3}
\end{eqnarray}
where $\alpha= \pm$, the $g^{\alpha}_{r}(k)$ form factors with $r=1$ to $6$ are defined in 
Fig. \ref{form-factors}, $E_l(\vec{k})$ 
are the eigenvalues of $M_k$ and $U_k$ is the matrix of eigenvectors. The $4$x$4$ matrices
$O_{X}(\bk)$ are defined by the schematic relation $ X =\frac{1}{N} \sum_{k}  \avg{ \h{\Psi}_k O_{X}(\bk) \Psi_k}$.
The bosonic distribution $n_b(x)$ is defined as
 $n_b(x) =  \sqrt{\frac{U}{2x}}\coth{\left(\beta\sqrt{\frac{U x}{2}}\right)}$ and $n_f(x)$ is the Fermi function.
The constraint equations for the number of fermions and bosons are given by
\begin{eqnarray}
\frac{1}{N} 
\sum_{k} \sum_{l=1}^4  n_f(E_{l}(\bk)) [\h{U}_k O_{n_s}(\bk) U_k]_{ll} & = & 1 ,\nonumber\\
 |u^{-}_{s}(0)|^2|z|^2 +  
\frac{1}{N}\sum_{k\neq 0,\alpha}
 |u_{s}^{\alpha}(\bk)|^2n_b(\gamma^{\alpha}_k)& = & 1, 
 \label{con2}
\end{eqnarray}
where $s=c,d$, the form factors $u^{\pm}_{s}(\bk)$ are the components of the rows of the transformation matrix that takes the
rotor action to diagonal form and $n_c$, $n_d$ are the densities on each sublattice. When the condensation amplitude is finite we also have
$\gamma^{-}_0 |z| = 0$. 

\begin{figure}
\includegraphics{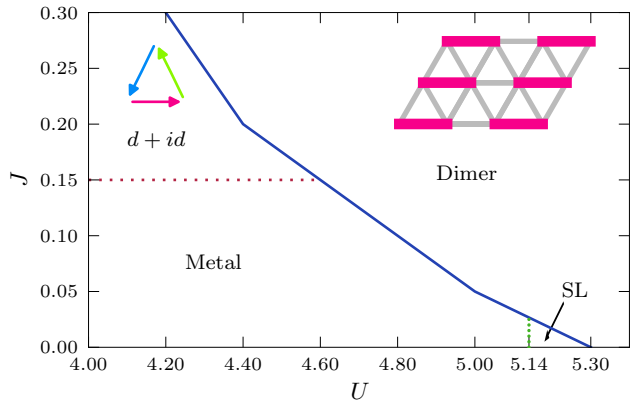}
\caption{
\label{zero-temp}
(Color online) The $J$ vs $U$ phase diagram. The dotted line
denotes a second-order transition, while the solid line is first-order.
The label $d+id$ denotes the superconducting phase, and the inset shows
the phase of $\Delta$ on each bond. The $U(1)$ spin liquid is labelled ${\rm SL}$. The pattern of $\chi$ and $B$ in
the dimer state is inset in the dimer phase, with thick lines denoting large
values and thin lines small values. We have rescaled $U \rightarrow U/2$ as discussed in \cite{rotor-florens-mott}.
}
\end{figure}

Fig. \ref{zero-temp} shows zero temperature phase diagram varying $J$ and $U$ obtained by solving the above coupled mean field equations Eq. \ref{mf3}
self-consistently while enforcing the constraints, Eq. \ref{con2}.
This involves 12 complex mean fields, 6 $\chi_{ij}$ and 6 $B_{ij}$ 
as indicated in Fig. \ref{unit-cell-fig}, linking the 
fermionic and bosonic mean field theories of the spinon and rotor sectors respectively.
For the pairing we use an ansatz of $\Delta_{r} = |\Delta|e^{i\theta_{\delta_r}}$ 
with uniform magnitude but allow the phase to vary within the unit cell. 

Four phases are identified in Fig. \ref{zero-temp}: a superconductor, a metal,
a spin liquid and the dimer state. 
The uniform metallic state is characterized by the condensation of
the bosonic degree of freedom and a uniform, real $\chi_{ij} = |\chi|$ 
implying the existence of an electron Fermi surface and gapless charge excitations.
On the insulating side, the spin liquid is also characterized 
by a uniform, real $\chi_{ij}$ but gapped bosons, meaning
the existence of a spinon Fermi surface but gapped charge excitations.
The dimerized phase is characterized by $\chi_{2} = 1/2$ and other $\chi_i=0$
(with a similar structure for $B$) and $\Delta=0$. Since the mean field solution consists of completely isolated
bonds, any dimer covering of the triangular lattice will have the same mean 
field energy at this level of approximation. 
In the superconductor we find the pairing is $d \pm id$,  as the phase winds by $2\pi/3$ around 
a triangle.
It is fully gapped, with no nodes. 
Note that we do not find a $\mathbb{Z}_2$ spin liquid, that is a spin liquid 
phase with finite spinon pairing but uncondensed rotor bosons.

One can show via perturbation theory that if $U/t \gg 1$, $B_{ij} \sim 4t \cc{\chi}_{ij}/U$, so
the spinon Hamiltonian becomes identical to a slave-fermion description
of the Heisenberg model with $J = 4t^2/U$. The result of
Rokhsar \cite{slave-fermion-rokhsar} states that under our conditions a dimerized
configuration is among the mean-field ground states. Thus we expect a dimerized
state at large $U$.  
However, the $U(1)$ spin liquid survives between the metal and dimer phases near MI transition
due to charge fluctuations.
The window of the spin liquid phase is quickly wiped
out as we increase $J$. One expects that as $J$ becomes large our mean field solution will be equivalent to
that of a Heisenberg model decoupled in $\chi$ and $\Delta$ channels. 
The argument of Rokhsar \cite{slave-fermion-rokhsar} can be generalized to this case,
and implies that the mean field ground states include dimerized states, 
where $\chi_{ij}=\Delta_{ij}=0$ on some bonds while on others $|\chi_{ij}|^2 + |\Delta_{ij}|^2$ is maximal.
Thus we expect that the dimerized state will be among the ground states
as we increase $J$.
Lee and Lee \cite{rotor-lee-lee} have looked at the pure Hubbard model ($J$=$0$ in our model) within the same mean field 
framework discussed above. However, they considered only translationally invariant solutions, 
and found a continuous transition from a metal into a spin liquid followed by 
a first order transition into a $\pi$-flux state. We find the $\pi$-flux state is replaced by the dimer phase,
and the window of spin liquid is also narrower than that of Ref.\cite{rotor-lee-lee}.

\begin{figure}
\centering
\includegraphics{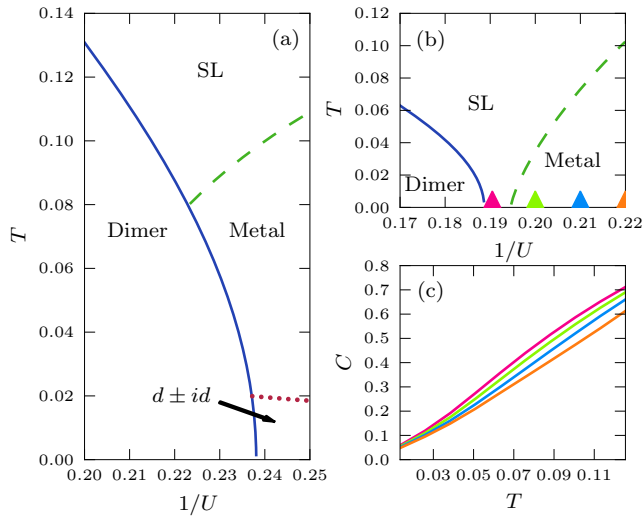}
\caption{
\label{finite-temp}
(Color online) Finite temperature phase diagrams, with $J=0.25$ for (a) and $J=0.00$ for (b). Curves have been fitted to the boundaries to guide the eye.
A dotted line denotes a second-order transition, a solid line is first-order and a dashed line is a crossover. Note the superconducting
phase at small $U$ for $J=0.25$ in (a). In (c) specific heat as a function of $T$ is shown for various values of $U$ denoted by
different colors in (b). Note that the slope of the curves increases with $U$ indicating more density of states in spin liquid insulator than metal.
 }
\end{figure}

It is straightforward to compute finite temperature phase diagram as shown in Fig. \ref{finite-temp} (a) and (b).
One peculiarity is that there is no condensation of the bosons in two dimensions at any finite temperature, and thus
no metallic phase. This is an artifact of the spherical approximation, that is treating the rotors as bosons \cite{rotor-florens-mott}. 
As the rotor action is equivalent to one used to describe Josephson junction arrays, we expect a
Kosterlitz-Thouless type transition from metal to spin liquid, with
quasi-long range order at low temperatures. In fact, including gauge fluctuations, the nature of the
phase transition from metal to spin liquid may be a crossover \cite{spin-liq-metal-ins,podolsky}. 
In Fig. \ref{finite-temp} (a) and (b) we  include a weak coupling between the layers via $-2 t_z {\rm cos}{k_z}$ in the dispersion with $t_z \sim 10^{-4}$, 
which should be present in quasi two-dimensional organic salts, to provide a consistent bosonic description for the whole phase space.

The $J=0.25$ phase diagram in Fig. \ref{finite-temp} (a) show a strong resemblance to  the phase diagram of $\EtMeP$ found in Refs. \cite{et-2007,et-2009}. 
Identifying the VB ordered state with the dimer state deserves some discussion.
First, since the dimer states are completely degenerate, any small anisotropy should break the degeneracy
and pick out a single dimer configuration, resulting in the VB ordered
phase seen in Ref. \cite{et-2006}. A coupling to the lattice should play an important role in this context
as discussed in \cite{et-2007}.  The spin susceptibility shows the expected
behavior as well, being exponentially suppressed in the dimer phase. Second, the $T_1^{-1}$ data for
the VB ordered state indicates a significant amount of inhomogeneity, possibly due to disorder. Taking
this disorder into account could drive the first order transition we have found theoretically to the
second order transition reported in \cite{et-2007}.  As well, one expects that including corrections
beyond our mean field theory could be  qualitatively similar to a quantum dimer model \cite{dimer-kivelson,dimer-read}.
It has been argued \cite{dimer-kt} that a VB solid phase in these models could show a second order transition
to a VB liquid phase at finite temperature, analogous to our dimer to spin liquid transition becoming second order.
At higher temperatures, the spin liquid reappears because of the large entropy associated with
the spinon Fermi surface and the gapped nature of the dimer state. Note that
near $U \sim 4.2 - 4.6$ as temperature increases the transitions, insulator (dimer) to metal
to insulator (spin liquid) occurs, similar to the re-entrance seen by resistivity experiments \cite{et-2007}.

At lower values of $J$ the spin liquid state persists even at zero-temperature, a scenario which has
been discussed extensively for the organic compound $\kappa-({\rm ET})_2{\rm Cu}_2({\rm CN})_3$ \cite{organic-kanoda-1,rotor-lee-lee,organic-motrunich}. The $J=0$ finite
temperature phase diagram is shown in Fig. \ref{finite-temp} (b) and has a window of spin liquid
between a metallic and dimer phase, which gets wider at finite temperature due to the larger entropy contained
in the spin liquid than the dimer phase.
It is tempting to argue that $\EtMeSb$ falls into the spin liquid regime, while
$\EtMeP$ belongs to the dimer phase nearby the spin liquid due to different exchange strengths $J$.
As we discussed, the gauge fluctuations are expected 
to turn the metal-spin liquid transition into a crossover rather than a sharp transition, but the details are still under debate. Due to
the paramagnetic state appearing above the VB ordered phase at temperatures comparable to this crossover
scale, reliable information on the gauge theory is needed to extract any experimentally relevant information.
However, we find that even at the level of mean field theory, there is difference in temperature dependence of 
the density of states in spin liquid and metal phases. The specific heat is plotted in spin liquid and metal in Fig. \ref{finite-temp} (c).
We note the linear behavior in both metal and insulating phase, with increasing slope as we pass from metal into
the spin liquid. This indicates a larger density of states at the Fermi level in the insulator, in contrast
to conventional MI transition where the density of states decreases to zero.

It is worthwhile to comment the absence of magnetic ordering in the phase space considered here. 
At large $U$ or $J$ one expects magnetic ordering of $120^{\circ}$ rotated spins. 
The magnetic order was considered in \cite{rotor-zhao}
where  the spin-spin term is decoupled into the $\chi$ and magnetic channels, imposing a $120^{\circ}$ degree
magnetic ordering. Although they treated the rotor sector differently, at
large $U/t$ both methods give the same asymptotic relation $B_{ij} \sim 4t\cc{\chi_{ij}}/U$,
which allows us to estimate the critical $U/t \approx 18$, above which the magnetic order sets in for $J/t=0.25$.
Since we are only interested in the intermediate $U$, our main results  near the MI transition are 
not affected by magnetic ordering of this type. Note that the tendency to order magnetically is also affected by
the anisotropy as we discussed above, which distinguishes EtMe$_3$ P and EtMe$_3$Sb from other $\dmit$ based
organic salts.

In summary, we propose an effective model for  the isotropic triangular ${\rm Pd}({\rm dmit})_2$ salts. 
We show that the Hubbard $U$ and spin exchange interactions account for the diversity present in the phase diagram of
the $\EtMeP$ salt, specifically the proximity of superconductivity and VB order near the pressure
induced MI transition. The spin liquid is realized for weak spin-spin interaction, which
may be relevant for the $\EtMeSb$ salt.  Our finite temperature phase diagram shows the re-appearance of the
spin liquid at finite temperatures due to entropic reasons, explaining the re-entrant insulator-metal-insulator transitions. We also present
the temperature dependence of the specific heat, which is qualitatively different between the spin liquid and metal and
 can be tested in future experiments.

\begin{acknowledgments}
We thank M. Hermele, Y.B. Kim, A. Paramekanti and S.S. Lee for useful discussions.
This research was supported by NSERC of Canada and the Canada Research Chair program.
\end{acknowledgments}


\end{document}